\documentclass[
reprint,
showpacs,preprintnumbers,
amsmath,amssymb,
aps,
pre,
]{revtex4-1}

\usepackage{graphicx}
\usepackage{dcolumn}
\usepackage{setspace}
\usepackage[utf8]{inputenc}
\usepackage{gensymb}
\usepackage{acronym}
\usepackage{hyperref}
\usepackage{xcolor}
\usepackage[percent]{overpic}

\newcommand\ie{i.\,e.}
\newcommand\psiglob{\Psi_{4}}
\newcommand\psiloc{\psi_4^{(n)}}
\newcommand\fig[1]{Fig.~\ref{#1}}
\newcommand\eq[1]{Eq.~\eqref{#1}}
\newcommand\eqs[1]{Eqs.~\eqref{#1}}
\newcommand\ourOnlyTable{Tab.~\ref{tab:c-exp}}
\newcommand\subfig[2]{Fig.~\ref{#1}(#2)}
\newcommand\subfigs[2]{Figs.~\ref{#1}(#2)}
\newcommand\subfiglong[2]{Figure~\ref{#1}(#2)}

\renewcommand\vec[1]{\boldsymbol{#1}}

\newcommand\rvec{\vec r}
\newcommand\kvec{\vec k}
\newcommand\rmi[2]{\ensuremath{#1_\text{#2}}}
\newcommand\bracket[1]{[#1]}
\newcommand\suscep{\chi_4}
\newcommand\rred{\epsilon}
\newcommand\rhoc{\rho_4}
\newcommand\rhopos{\rmi \rho {pos}}
\newcommand{\tightoverset}[2]{\mathop{#2}\limits^{\vbox to -.5ex{\kern-1.2ex\hbox{$#1$}\vss}}}
\newcommand\com[1]{\ensuremath{#1_{\mathrm{com}}}}

\newcommand\colwidth{0.44}
\newcommand{\coll}{c}
\newcommand{\fluid}{d}
\newcommand{\round}{e}
\newcommand{\patchy}{f}
\newcommand{\four}{g}
\newcommand{\triang}{h}
\newcommand{\bulky}{i}
\newcommand{\band}{j}

\newacro{ness}[NESS]{nonequilibrium steady state}
\newacroplural{ness}[NESSs]{nonequilibrium steady states}
\newacro{md}[MD]{molecular dynamics}
\newacro{2d}[2D]{Two-dimensional}
\newacro{bkthny}[BKTHNY]{Berezinskii-Kosterlitz-Thouless-Halperin-Nelson-Young}

\begin{document}

\title{Nonequilibrium steady states, coexistence and criticality
	\\in driven quasi-two-dimensional granular matter}

\author{Thomas Schindler}
\email{Thomas.Schindler@fau.de}
\author{Sebastian C.\ Kapfer}
\email{Sebastian.Kapfer@fau.de}
\affiliation{Theoretische Physik 1, FAU Erlangen-N\"urnberg, Staudtstr.~7, 91058 Erlangen, Germany}

\date{\today}

\begin{abstract}
Nonequilibrium steady states of vibrated inelastic frictionless spheres are investigated in quasi-two-dimensional confinement via molecular dynamics simulations.
The phase diagram in the density-amplitude plane exhibits a fluidlike disordered and an ordered phase with threefold symmetry, as well as phase coexistence between the two.
A dynamical mechanism exists that brings about metastable traveling clusters and at the same time stable clusters with anisotropic shapes at low vibration amplitude.
Moreover, there is a square bilayer state which is connected to the fluid by BKTHNY-type two-step melting with an intermediate tetratic phase.
The critical behavior of the two continuous transitions is studied in detail.
For the fluid-tetratic transition, critical exponents of $\tilde{\gamma}=1.73$, $\eta_4 \approx 1/4$, and $z=2.05$ are obtained.
\end{abstract}

\maketitle

\section{Introduction}\label{sec:intro}

The slab geometry of vibrating plates filled with inelastic granular spheres is a particularly interesting setup in the field of nonequilibrium statistical physics.
The macroscopic particles (typical size $\sim 1\,\mathrm{mm}$) dissipate kinetic energy at each collision.
Fluidized states can be maintained by energy input to the vertical particle motion via the vibrating plates.
This energy is then partly transferred to horizontal motion in particle-particle collisions.
The energy flow in this injection, transfer, and dissipation mechanism breaks detailed balance and the system is inherently strongly out of equilibrium.
The nonequilibrium property manifests itself in several interesting phenomena which have been investigated largely in the past two decades.
Among these are
inelastic collapse~\cite{OlafsenPRL1998,NieEPL2000,OlafsenPRL2005,KhainPRE2011},
inhomogeneous granular temperatures~\cite{PrevostPRE2004,LobkovskyEPJST2009},
non-Gaussian velocity distributions~\cite{LosertCHAOS1999,OlafsenPRE1999,KawaradaJPSJ2004},
segregation of mixtures~\cite{RivasPRL2011,RivasNJP2011,RivasGM2012},
the Kovacs memory effect~\cite{PradosPRL2014,TrizacPRE2014,BreyPRE2014},
and inelastic hydrodynamic modes~\cite{BritoPRE2013}.

\ac{2d} driven granular matter excels
as a model system for nonequilibrium statistical mechanics for several reasons.
Particle trajectories are comfortably accessible in experiments by filming from the top.
Moreover, the influence of gravity is tunable via the choice of the driving amplitude and frequency~\cite{MelbyJPCM2005}.
Finally, the particles can be agitated homogeneously throughout the horizontal directions.
Any in-plane inhomogeneity thus emerges from spontaneous symmetry breaking~\cite{MoonPRE2004}.

After some relaxation time, a \ac{ness} is reached in which the energy injection, transfer, and dissipation rates balance.
Because of their nonequilibrium nature, the involved phases in these states are no thermodynamic phases in the strict sense.
Even though these phases show intriguing resemblance to the corresponding equilibrium system~\cite{SchmidtPRE1997} of colloidal particles, the \acp{ness} usually retain residual energy and particle flows which cannot occur in equilibrium.

The phase behavior depends decisively on parameters such as box dimensions and roughness, filling density, driving frequency and amplitude, and inelasticity of the particles.
Several studies have investigated the phase behavior as functions of different control parameters~\cite{PrevostPRE2004,MelbyJPCM2005,ReisPRL2006,ClercNP2008,VegaPRE2008,RivasNJP2011,GuzmanPRE2018}.
A complete phase diagram in the multi-dimensional parameter space, however, is not at hand.
For the parameters studied here, the \acp{ness}
comprise isotropic fluid-like phases as well as square and hexagonal monolayers and bilayers.
With equilibrium hard spheres, the transitions between those phases are all of first-order type~\cite{SchmidtPRE1997}.
By contrast, in experiments with the shaken granular particles, a continuous transition between an isotropic phase and a phase with square order has been reported, with diverging correlation functions and several critical exponents measured~\cite{CastilloPRL2012,CastilloPRE2015}.
In \ac{md} computer simulations the same phases were found~\cite{GuzmanPRE2018}, but divergences of the correlation functions were not reproduced.

The fact that -- despite extensive studies -- the picture of the quasi-\ac{2d} vibrated granulates is still incomplete led us to revisit the system with \ac{md} simulations.
We employ an altered approach to refine the description of the continuous fluid-square transition and characterize it as \ac{bkthny} type~\cite{BerezinskiiJETP1971,KosterlitzJPC1974,HalperinPRL1978,YoungPRB1979} two-step transition with an intermediate tetratic phase.
The tetratic phase is characterized by (quasi-)long-ranged orientational but short-ranged positional order, and has also been found in equilibrium quasi-\ac{2d} Hertzian spheres \cite{TeraoJCP2013}.
Moreover, a nonequilibrium phenomenon is described, namely, emergent particle currents at the surfaces of threefold clusters which can lead to macroscopic cluster propulsion.

The paper is organized as follows.
In Sec.~\ref{sec:sys} we present the system and model and give technical details about the simulation and the order parameter.
Secs.~\ref{sub:diag}--\ref{sub:facet} summarize our main findings on the \acp{ness}.
First, we discuss the relevant parameters for the formation of ordered phases and sketch the \ac{ness} phase diagram in Sec.~\ref{sub:diag}.
Second, the fluid-tetratic-square transition contained in the phase diagram is thoroughly characterized in Sec.~\ref{sub:crit},
including precise values for the critical exponents.
Finally, the mechanism that brings about anisotropic shape and persistent motion of threefold clusters is discussed in Sec.~\ref{sub:facet}.
In Sec.~\ref{sec:con} we conclude by comparing the phase behavior to previous studies and discuss the nonequilibrium nature of the observed effects.

\section{System setup and parameters}\label{sec:sys}

Computer simulations are performed in a shallow cuboidal box of spatial dimensions $L\times L\times h$ with periodic boundary conditions applied in $x$ and $y$ directions.
In $z$ direction the box is confined by two hard walls.
The space in between the walls contains $N$ hard spheres of diameter $\sigma$ at a projected number density $\rho\equiv N/L^2$.
Gravitational acceleration $g$ acts on the particles in negative $z$ direction implying a timescale $\tau_0\equiv \sqrt{\sigma/g}$.
The walls oscillate in-phase with displacement $A\sin(\omega t)$ in $z$ direction, where $A$ is the driving amplitude, $\omega$ is the angular frequency of the driving, and $t$ denotes time.
All simulations in this work are carried out with $h=1.83\sigma$ and $\omega\tau_0=12$.
The height was chosen such that fluid, threefold, and square phases compete.
(The phases are described in detail in Sec.~\ref{sub:diag}.)
The value of $\omega$ lies within the high-frequency regime~\cite{PrevostPRE2004}, where the nucleation of the threefold cluster upon increasing $A$ is insensitive to $\omega$.

The trajectories of the particles are calculated by an event-driven \ac{md} algorithm~\cite{BannermanJCC2011}.
This algorithm is appropriate for instantaneous interaction events (collisions).
In the time intervals between events, particles move on ballistic trajectories. Therefore, collision times can be calculated from the initial configuration (positions and velocities) and scheduled chronologically.
The main loop then evolves the system in time by processing the events in the schedule. In each event, the velocities of the collision partners are changed and the next collision times of these particles are calculated and inserted into the schedule.
Particle-particle and particle-wall collisions are modeled as inelastic collisions~\cite{BrilliantovPoeschel} with a constant coefficient of restitution $e=0.95$ for the  momentum transfer normal to the surfaces (where $e=1$ and $e=0$ would correspond to elastic and fully inelastic collisions, respectively).

The model conserves momentum in $x$ and $y$ directions (momentum in $z$ direction changes in particle-wall collisions).
We do not include transfer of momentum tangential to the particle surfaces, \ie, the Coulomb friction coefficient vanishes.
Hence there is no coupling of rotational and translational degrees of freedom.
Our model thus depends only on a single parameter -- the coefficient of restitution -- yet still features energy injection, transfer, and dissipation mechanisms.
Despite this simplification, our phase diagram qualitatively agrees with earlier studies that do include the rotational degrees of freedom.

\begin{figure}
	\includegraphics[width=0.27\textwidth]{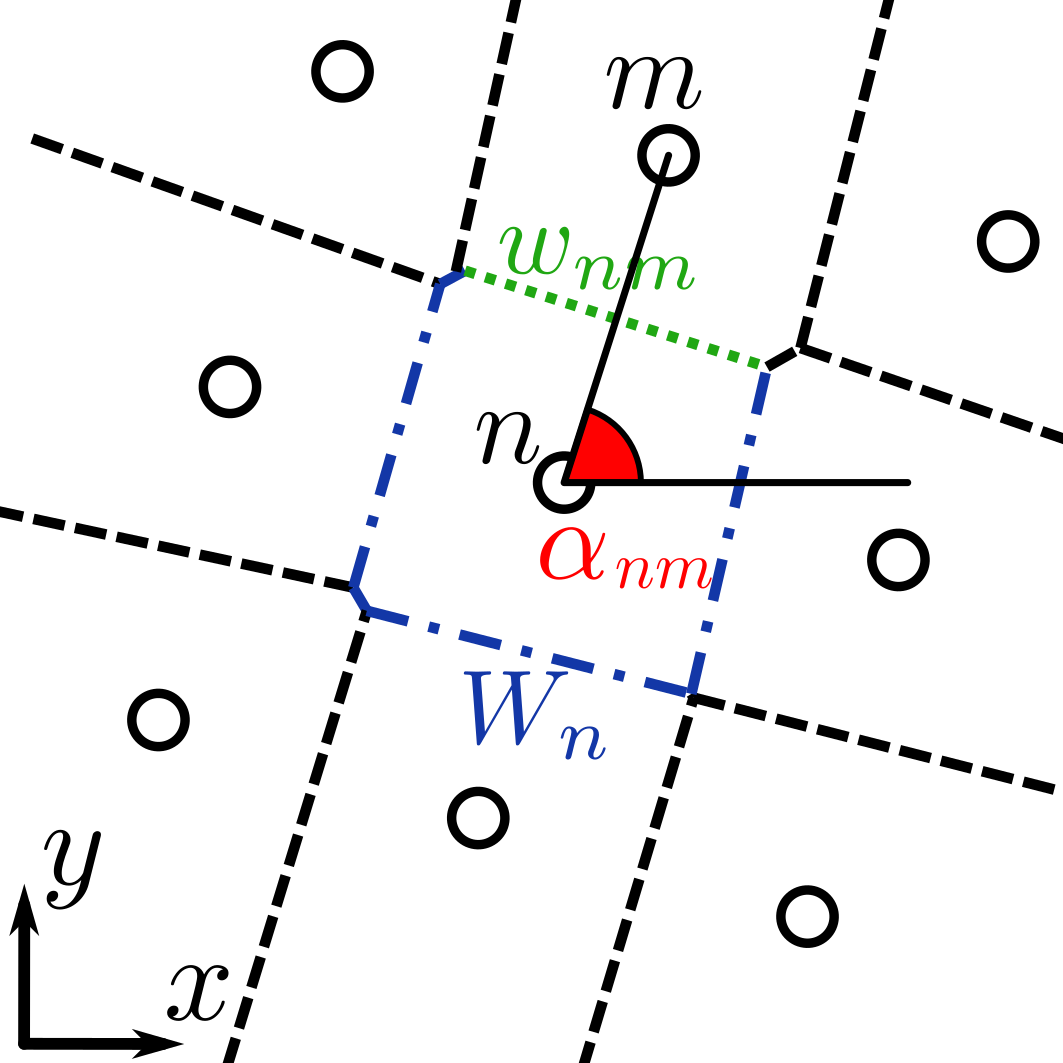}
	\caption{\label{fig:Voronoi}(color online)
		Sketch of geometrical objects defining the \ac{2d} local bond-orientational order parameter $\psiloc$.
		Particle centers are depicted as circles.
		Facets of the Voronoi tessellation (\ie, the perpendicular bisections of the connections of particle centers) are drawn as dashed, dotted, and dashed-dotted lines.
		Particles that share a Voronoi facet are considered nearest neighbors.
		Symbols and coloring of the sketch are described in the text.
	}
\end{figure}

Most of our analysis and the color coding of all snapshots in this work are based on the (projected \ac{2d}) fourfold local bond-orientational order parameter $\psiloc$ of a particle $n$.
We use a refined version~\cite{MickelJCP2013},
\begin{equation}
\psiloc\equiv\sum_{m=1}^{N_n}\frac{w_{nm}}{W_n}\mathrm e^{4\mathrm i\alpha_{nm}},
\end{equation}
with weight factors differing from the usual definition.
Here the sum is carried out over the $N_n$ Voronoi nearest neighbors of particle $n$, the weight factor $w_{nm}/W_n$ is the length $w_{nm}$ of the Voronoi facet shared by particles $n$ and $m$ (green dotted line in \fig{fig:Voronoi}) normalized by the total perimeter $W_n$ of the Voronoi cell of particle $n$ (green dotted plus blue dashed-dotted lines), and $\alpha_{nm}$ (red) is the angle between the $x$ axis and the connection line of particles $n$ and $m$.
The weight factors are included to make $\psiloc$ a continuous function of particle positions and in particular robust against small distortions in lattices by minimizing the influence of e.\,g.\ diagonal nearest neighbors in distorted square lattices (like the top left particle in \fig{fig:Voronoi}).
The modulus of $\psiloc$ ranges from 0 for particles with no local fourfold symmetry to 1 for particles centered in a square of four nearest neighbors.

Unless otherwise stated, each simulation run is started from a special configuration
prepared to minimize nucleation effects.
The configuration consists of a domain of square order with local density of $1.6\sigma^{-2}$ and a domain of threefold order (local density $2.25\sigma^{-2}$),
immersed in a fluid (local density $0.8\sigma^{-2}$).
Velocities are initialized from Gaussian distributions, where the mean velocity of the $N$ particles is subtracted yielding zero net velocity.
After a relaxation period the simulations reach steady states which are the subjects of our investigations.
We explicitly verified that all steady states are stable when the simulation is paused and the velocities are reset
to a Gaussian distribution.

\section{Phase diagram}\label{sub:diag}

\begin{figure*}
	\begin{overpic}[width=\textwidth]{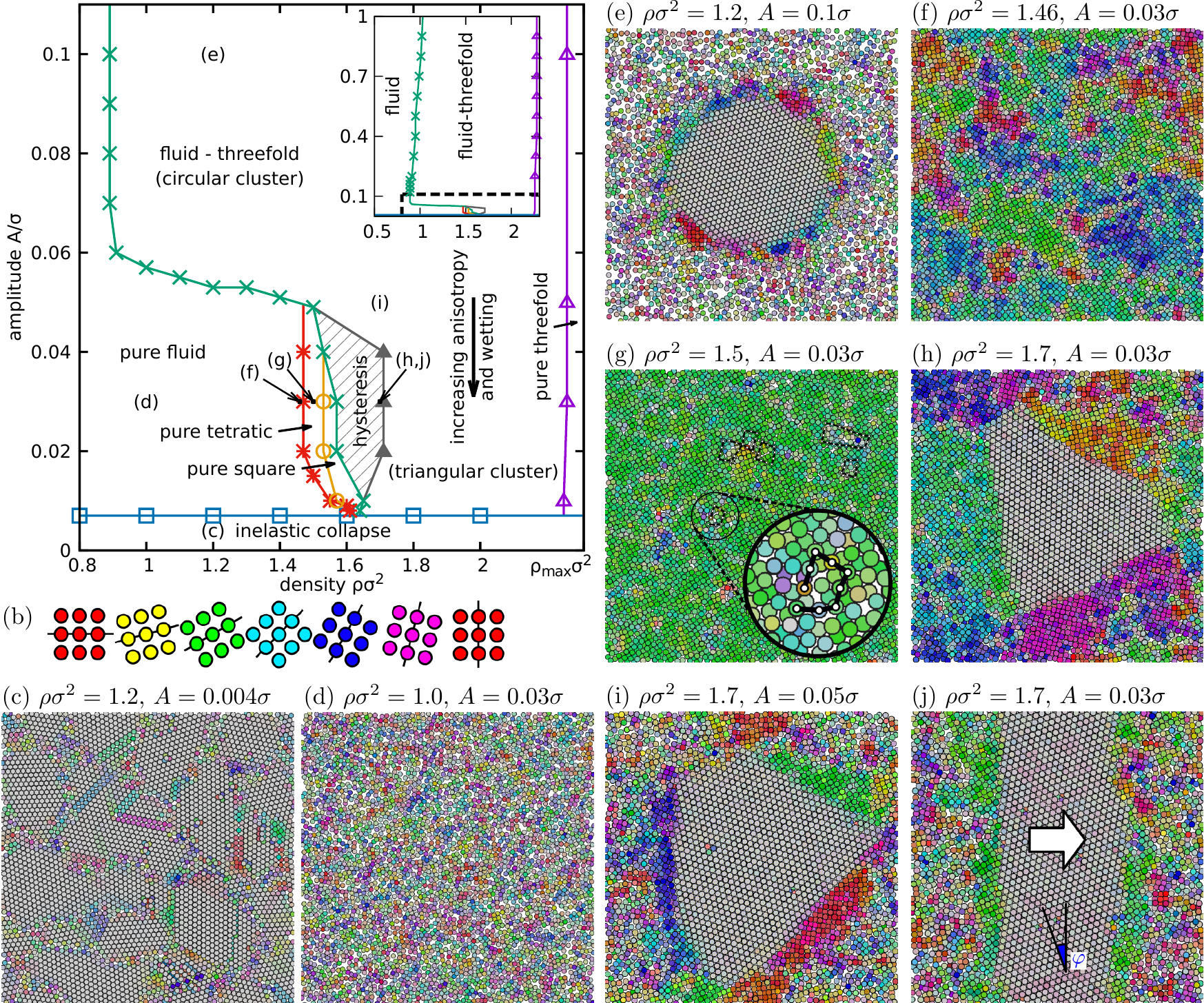}
		\put (0.2,81.6) {(a)}
	\end{overpic}
    \caption{\label{fig:pd}(color online)
		\textbf{(a)} Phase diagram in the density-amplitude plane
			for box height $h=1.83\sigma$ and angular frequency $\omega\tau_0=12$.
			Investigated state points are on a grid with spacings $\Delta \rho\sigma^2=0.02$ and $\Delta A=0.002\sigma$.
			The simulation at each state point was carried out with $N=4000$ particles and averaged over a time interval $t=5000\tau_0$.
			The maximum density $\rmi \rho {max} \sigma^2 \approx 2.31$ is  slightly larger than the density of two hexagonal close packed layers, $\rho\sigma^2=4/\sqrt{3}$ due to the possibility of buckling.
			Below the blue squares the fluidized state collapses and all particles drop to the bottom plate.
			The green crosses indicate evaporation of a threefold cluster upon decreasing $\rho$.
			Between $0.02\sigma \leq A \leq 0.04\sigma$ the nucleation density (gray filled triangles) of the threefold cluster upon compression differs from the evaporation density, with a hysteresis region (hatched) between the green crosses and gray filled triangles.
			The purple open triangles mark the density where the threefold cluster comprises the whole simulation box.
			The red stars and yellow circles indicate the continuous fluid-tetratic and tetratic-square transitions, respectively.
			Labels (c)--(j) relate the state points to the snapshots in the following panels.
			Lines are guide to the eye.
			The inset shows a wider view of the phase diagram up to amplitude $A=1\sigma$ (where the dashed line marks the view of the main panel).
		\textbf{(b)} Sketch of the color coding of the particles in snapshots (c)--(j).
			The hue is determined by the complex phase $\arg(\psiloc)$ of the local order parameter.
			The color saturation indicates the modulus of $\psiloc$ ($|\psiloc|=0$ $\rightarrow$ gray, $|\psiloc|=1$ $\rightarrow$ fully saturated).
		(c)--(j) Top view snapshots of the simulation box at state points indicated and labeled in the phase diagram:
		\textbf{(\coll)} $\rho\sigma^2=1.2$, $A=0.004\sigma$ inelastic collapse (nonergodic).
		\textbf{(\fluid)} $\rho\sigma^2=1.0$, $A=0.03\sigma$ fluid phase.
		\textbf{(\round)} $\rho\sigma^2=1.2$, $A=0.1\sigma$ fluid-threefold coexistence (circular cluster shape).
		\textbf{(\patchy)} $\rho\sigma^2=1.46$, $A=0.03\sigma$ fluid with square bilayer patches.
		\textbf{(\four)} $\rho\sigma^2=1.5$, $A=0.03\sigma$ tetratic phase.
			The magnifier shows a connected path along grid lines, which follows two lattice sites up, two right, two down, and two left.
			The fact that the path does not close, as it would in a regular lattice, indicates a dislocation defect.
			The other dislocations are marked in the same way.
		\textbf{(\triang)} $\rho\sigma^2=1.7$, $A=0.03\sigma$ fluid-threefold coexistence (with the cluster wetted by square bilayer phase; triangular cluster shape).
		\textbf{(\bulky)} $\rho\sigma^2=1.7$, $A=0.05\sigma$ fluid-threefold coexistence (partially wetted).
		\textbf{(\band)} $\rho\sigma^2=1.7$, $A=0.03\sigma$ fluid-threefold coexistence with cluster percolating the box in $y$ direction \bracket{metastable; same state point as (h)}.
		$\varphi$ denotes the angle between the $y$ axis and the symmetry axis of the lattice unit cell (see Sec.~\ref{sub:facet}).
		The big arrow indicates the direction of motion of the cluster.
	}
\end{figure*}

Various aspects of the phase diagram have been reported previously~\cite{OlafsenPRL1998,PrevostPRE2004,MelbyJPCM2005,VegaPRE2008,LobkovskyEPJST2009,GuzmanPRE2018}, but a complete picture is lacking so far.
In \subfig{fig:pd}{a} we present the phase diagram in the $\rho$-$A$ plane, obtained for $N=4000$ particles.
(The transition lines shown here are shifted with respect to the true ones by finite-size effects.)
For the chosen parameters, the system exhibits both transitions with first-order
character as well as critical behavior.

At low driving amplitudes $A<0.007\sigma$,
we observe inelastic collapse~\cite{OlafsenPRL1998,BrilliantovPoeschel}
in which all spheres drop to the bottom plate because the injected energy does not suffice to maintain a fluidized state.
A snapshot of this state is displayed in \subfig{fig:pd}{\coll}; as in all snapshots, the colors show the local orientation of the square order, \ie, the complex phase of $\psiloc$, as depicted in \subfig{fig:pd}{b}.

At higher $A$ there are several fluidized states described in the following.
At low $\rho$, the system is in a homogeneous unordered fluid phase, exemplarily seen in \subfig{fig:pd}{\fluid}.
In the high-density limit, we find a lattice with threefold symmetry.
This lattice consists of two hexagonal layers offset against each other, such
that particles of the top layer sit in the dips of the bottom layer, as in the
hexagonal close-packed structure.
In projection, one finds a honeycomb lattice with three nearest neighbors.
Because bottom and top layer particles are distinct in the gravitational field, this lattice only has threefold rotational symmetry and not sixfold as the honeycomb (see also Sec.~\ref{sub:facet}).
The transition from the fluid to the threefold lattice is different in the high-amplitude regime $A\gtrsim0.05\sigma$ and in the moderate-amplitude regime $0.007\sigma < A \lesssim 0.05\sigma$.

In the high-amplitude regime the transition exhibits the phenomenology of a first-order phase transition.
At the evaporation density $\rho\sigma^2\approx0.9$, a cluster with threefold structure emerges \bracket{see \subfig{fig:pd}{\round}} which then grows with increasing $\rho$ until it comprises the entire box for $\rho\sigma^2\gtrsim2.26$.
This transition scenario is stable up to at least $A=1\sigma$ as shown in the inset of \subfig{fig:pd}{a}.
The relatively broad coexistence region as compared to the thermal equilibrium system~\cite{SchmidtPRE1997} is an effect of enhanced dissipation in the dense phase and has been reported previously~\cite{PrevostPRE2004,MelbyJPCM2005,VegaPRE2008}.

In the moderate-amplitude regime, the fluid contains patches with square bilayer structure \subfig{fig:pd}{\patchy} for $\rho\sigma^2\gtrsim 1.4$.
The length scale and life time of these patches (as seen in movie no.~1 in the Supplemental Material~\cite{Supp}) diverges upon increasing $\rho$, and the system undergoes a continuous transition.
The result is the tetratic state as seen in \subfig{fig:pd}{\four}.
This phase is distinguished from a true solid by the presence of dislocations \bracket{marked in \subfig{fig:pd}{\four}} at which grid lines end.
In a second continuous transition at higher $\rho$, the density of dislocations vanishes and
a square bilayer solid is formed (no snapshot shown).
This two-step transition hence displays the phenomenology of the \ac{bkthny} theory~\cite{HalperinPRL1978,YoungPRB1979} and is analyzed in greater detail in Sec.~\ref{sub:crit}.
In contrast to the large density difference between the fluid and the threefold phase,
we do not find any evidence of density inhomogeneities at the fluid-tetratic or tetratic-square transitions.

At higher $\rho$, a first-order-type transition to the threefold lattice is found with an evaporation density of the threefold cluster of $\rho\sigma^2\approx 1.57$.
Surprisingly, in the coexistence region, the square phase is destabilized by the presence of the threefold cluster and melts into a fluid, see \subfig{fig:pd}{\triang}.
This seemingly paradoxical topology of the phase diagram may either be a finite-size effect
due to the size of the critical nucleus
(though dilution persists in $N=16000$ simulations,
see movie no.~2 in the Supplemental Material~\cite{Supp}),
or a genuine nonequilibrium feature (see discussion).

From what we can tell from our simulations,
the fluid-tetratic-square transition and the density instability leading to clusters are independent.
In other words there is no extra critical behavior at the state points $A=0.05\sigma$, $\rho\sigma^2=1.47$ and $A=0.04\sigma$, $\rho\sigma^2=1.53$,
where the critical transition lines intersect with the evaporation line of the threefold cluster.

To check if the fluid-threefold coexistence is the true \ac{ness} above the evaporation density, additional simulations were performed in that region, initialized as pure square phases (including grain boundaries to facilitate nucleation) or fluids.
Threefold clusters are indeed nucleated, but at very low rate. In the range $0.02\sigma \leq A \leq 0.04\sigma$ and at $\rho\sigma^2\leq 1.71$, however, we could not observe any nucleation events at all, which is indicated in the phase diagram as a hatched hysteresis region.
The relative stability of the fluid-tetratic-square branch and the
demixed fluid-threefold state is therefore inconclusive. (In nonequilibrium,
we cannot determine the free energies of the two branches.)  The two Berezinskii-Kosterlitz-Thouless transitions
might thus lie on a metastable branch.

When avoiding nucleation, the transition lines of the fluid-tetratic-square transition continue metastably up to $A\approx0.068\sigma$. At lower $\rho$, we find coexistence of two fluids with distinct densities in the range $0.05\sigma \lesssim A \lesssim 0.068\sigma$ and fluid-square coexistence for $A\gtrsim 0.068\sigma$, qualitatively consistent with the findings of Guzm\'an \textit{et al.}~\cite{GuzmanPRE2018}.
These states are metastable in our simulations, however.

Finally, we discuss the shape of the threefold cluster which appears in the fluid-threefold coexistence.
For large $A$, the shape of the cluster is dominated by an isotropic surface tension and thus is close to circular, \subfig{fig:pd}{\round}.
At lower $A$ the symmetry breaking between the top and bottom hexagonal layers due to gravity is enhanced.
This leads to a pronounced anisotropy of the surface tension, and to the emergence
of three stable and three unstable directions in the hexagonal bilayer (see Sec.~\ref{sub:facet}).
The three stable facets grow out to become the three corners of a cluster with triangular shape \bracket{see \subfigs{fig:pd}{\triang} and (\bulky)}.
The remaining interfaces of this cluster are of the unstable facet type.

In finite (square) simulation boxes with periodic boundary conditions, phase coexistence manifests in three different topologies of clusters~\cite{MayerJCP1965}:
a threefold cluster surrounded by fluid [\subfigs{fig:pd}{\round), (\triang), and (\bulky}], a stripe-shaped cluster [\subfig{fig:pd}{\band}], and a drop of fluid within a threefold lattice (not shown).
A detailed investigation of the relative stability of these structures has not been done here.
The stripe geometry is of particular interest, however.
If the two fluid-solid interfaces are inequivalent, which is possible due to broken parity of the threefold lattice,
the cluster may absorb new particles on one interface \bracket{right-hand side in \subfig{fig:pd}{\band}}
and dissolve on the other (left), leading to the curious effect that the ordered domain is effectively propelled forward (see movie no.~3 in the Supplemental Material~\cite{Supp}).
The mechanism behind this and the dependence of the effective cluster speed on its orientation and $A$ is examined in Sec.~\ref{sub:facet}.

\section{Fluid-tetratic-square transition}\label{sub:crit}

The freezing of the fluid to the square solid proceeds in two continuous phase transitions at two different critical densities.
At the lower density, the fluid transforms into a tetratic state by divergence of the length and timescales of ordered patches.
In the following, we measure the emerging orientational order and calculate precise values for the fluid-tetratic critical density $\rhoc$ and the critical exponents governing the divergences.
The system is driven through the transition by increasing $\rho$ at several fixed $A$.

\begin{figure}
	\includegraphics[width=\colwidth\textwidth]{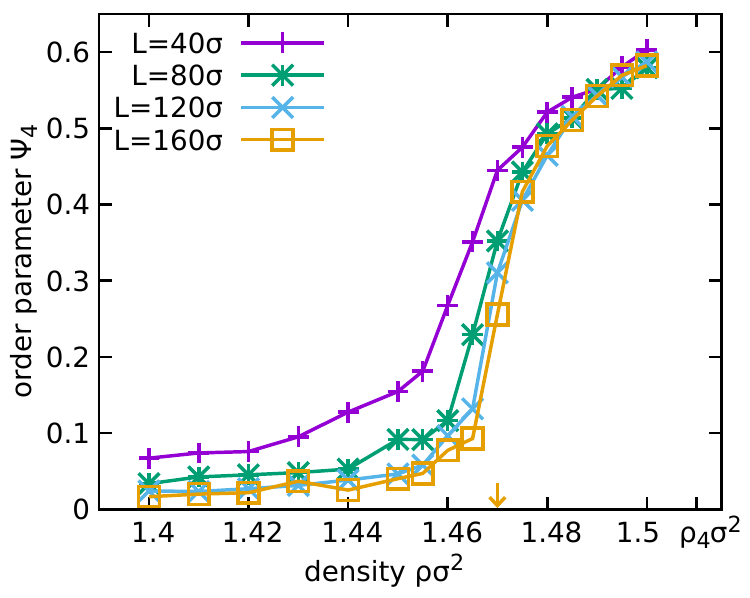}
	\caption{\label{fig:psi-glob}(color online)
		Order parameter $\psiglob$ as a function of density $\rho$
		for $A=0.03\sigma$ and different box sizes $L$ as indicated.
		Lines are guides to the eye.
		$\rhoc$ indicates the fluid-tetratic critical density as fitted via \eq{eq:xi-fit}. The arrow at $\rho\sigma^2=1.47$ marks the density where at $L=160\sigma$ crossover from short-ranged to quasi-long-ranged behavior is observed in the fourfold correlation function (cf.\ \fig{fig:g4}).
	}
\end{figure}

At first the global orientational order for several system sizes, ranging from $L=40\sigma$ to $160\sigma$, and $A=0.03\sigma$ is considered as an indicator for a continuous transition.
The degree of global orientation is measured via the order parameter
\begin{equation}
\label{eq:psi-glob-def}\psiglob\equiv \left\langle\left|\frac{1}{N}\sum_{n=1}^{N}\psiloc\right|\right\rangle,
\end{equation}
where the angle brackets denote time average. The typical behavior of a continuous transition is observed (cf.\ Fig.~\ref{fig:psi-glob}):
in the fluid phase, like the one displayed in \subfig{fig:pd}{\fluid}, contributions of the differently oriented particles cancel out, yielding zero mean.
When approaching the critical point, patches with fourfold orientational order emerge,
which increase in size and eventually reach the scale of the box.
In this region there are only a few patches \bracket{see, e.\,g., \subfig{fig:pd}{\patchy}}. Their contributions to $\psiglob$ are unlikely to cancel completely, as would be the case for many small patches, and therefore yield a finite average.
The effect is more prominent with smaller simulation boxes and sets in at lower $\rho$.
When further increasing $\rho$, there is only one domain left and the whole system orders, yielding a strong increase of $\psiglob$.
The resulting kink in the data is at a density lower than $\rhoc$, but it approaches $\rhoc$ in the limit of $L\rightarrow\infty$.
Note that because the tetratic phase does not exhibit true long-ranged orientational order,
an exponent $\beta$ governing the order parameter via $\psiglob\sim(-\epsilon)^\beta$
does not exist in the infinite system~\cite{BramwellJPCM1993}.

\begin{figure*}
	(a)\raisebox{\glueexpr\baselineskip -1\glueexpr\height}{\includegraphics[width=0.3\textwidth]{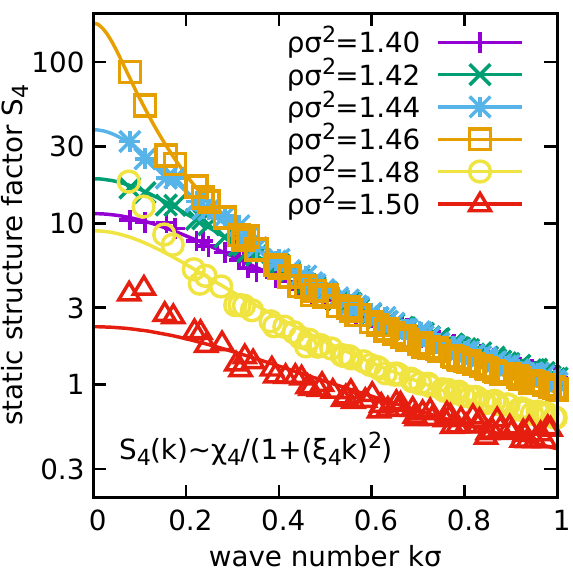}}
	(b)\raisebox{\glueexpr\baselineskip -1\glueexpr\height}{\includegraphics[width=0.3\textwidth]{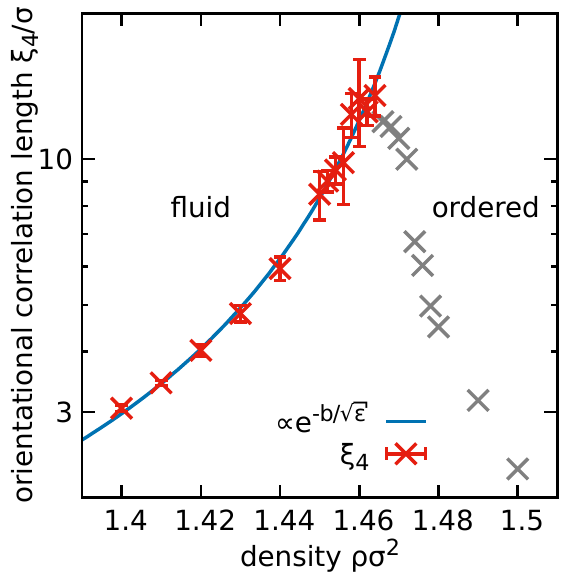}}
	(c)\raisebox{\glueexpr\baselineskip -1\glueexpr\height}{\includegraphics[width=0.3\textwidth]{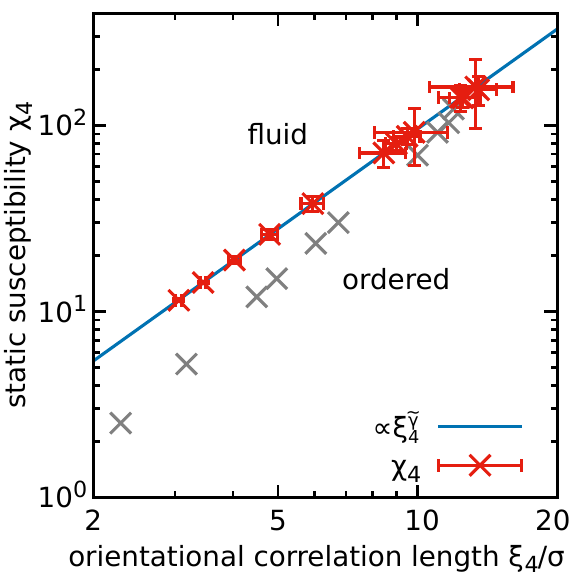}}
	\caption{\label{fig:stat-corr}(color online)
		(a) Static structure factor $S_4$ as a function of wave number $k$
			in the critical region exemplarily at densities $\rho$ indicated in the legend obtained at amplitude $A=0.03\sigma$
			and box size $L=120\sigma$.
			Lines are fits to the Ornstein-Zernike form, \eq{eq:OrnsteinZernike}, only valid while the finite system is fluid for $\rho\sigma^2\leq 1.464$.
		(b) Orientational correlation length $\xi_4$ as a function of density $\rho$ as extracted from the data in (a).
			The blue line is a fit to \eq{eq:xi-fit} using data of the fluid regime (red symbols with error bars).
			Gray symbols without error bars are ordered states.
		(c) Static susceptibility $\suscep$ as a function of $\xi_4$ as extracted from the data in (a).
			The blue line is a fit to \eq{eq:chi-fit} using data of the fluid regime (red symbols with error bars).
			Gray symbols without error bars are ordered states.
	}
\end{figure*}

Length and timescales are studied via
the (fourfold) intermediate scattering function~\cite{HMcD}
\begin{equation}\begin{split}
F_4(\kvec,\tau)\equiv\frac{1}{N}\left<\sum_{m=1}^{N}\sum_{n=1}^{N}\right.& \mathrm e^{\mathrm i\kvec\cdot(\rvec_m(t)-\rvec_n(t+\tau))}\\
&\psi_4^{(m)}(t)\bar{\psi}_4^{(n)}(t+\tau)\left.\vphantom{\sum_{m=1}^{N}\sum_{n=1}^{N}}\right>,
\end{split}\end{equation}
where $\kvec$ is a \ac{2d} wave vector, $\rvec_m$ is the \ac{2d} projection of the position of particle $m$, $\tau$ is a time difference, and the bar denotes complex conjugation.
At equal times, $F_4$ is the static structure factor, $S_4(\kvec)\equiv F_4(\kvec,0)$.

In the fluid, for $\rho<\rhoc$, Ornstein-Zernike behavior~\cite{HMcD}
\begin{equation}
\label{eq:OrnsteinZernike}S_4(\vec k)= \frac{\suscep}{1+(\xi_4\vec k)^2},
\end{equation}
at low wave numbers $k\equiv |\kvec|$ is observed.
We determine the static susceptibility $\suscep$ and the orientational correlation length $\xi_4$ \bracket{\ie, the typical size of a patch as seen in \subfig{fig:pd}{\patchy}}
by fitting our simulation data to \eq{eq:OrnsteinZernike}.
\subfiglong{fig:stat-corr}{a} shows $S_4(\vec k)$ for several densities and $L=120\sigma$ exemplarily at $A=0.03\sigma$.
Data for other values of $A$ ranging from $0.01\sigma$ to $0.04\sigma$ obtained with $L=80\sigma$ show qualitatively the same behavior.
The extracted values of $\xi_4$ are displayed in \subfig{fig:stat-corr}{b}.
The data exhibits divergence at $\rhoc$ of the $XY$ model type~\cite{KosterlitzJPC1974},
\begin{equation}
\xi_4\propto \exp(b/\sqrt{\rred}),\label{eq:xi-fit}
\end{equation}
with constant $b$ and reduced density parameter $\rred\equiv 1-\rho/\rhoc$.
The finite system already orders at lower $\rho$, where $\xi_4$ attains its maximum. Fits are therefore restricted to the region $\rho\sigma^2\leq 1.464$ (red data points with error bars).

The divergence of $\suscep$ is closely linked to the divergence of $\xi_4$, as shown in \subfig{fig:stat-corr}{c}.
One observes power law dependence
\begin{equation}
\suscep\propto \xi_4^{\tilde{\gamma}},\label{eq:chi-fit}
\end{equation}
with critical exponent $\tilde{\gamma}$.
The resulting values of $b$, $\rhoc$, and $\tilde{\gamma}$ at all investigated $A$ are displayed in \ourOnlyTable.

\begin{table*}
	\begin{tabular}{
			c
			| D{,}{\,\pm\,}{4,4}  D{,}{\,\pm\,}{5,5}
			| D{,}{\,\pm\,}{5,5}
			| D{,}{\,\pm\,}{5,5}
		}
		\hline\hline
		\vphantom{$X^{X^{X^{X}}}$}
		& \multicolumn{2}{c|}{$\xi_4\propto \exp(b/\sqrt{\rred})$}
		& \multicolumn{1}{c|}{$\suscep\propto \xi_4^{\tilde{\gamma}}$}
		& \multicolumn{1}{c}{$\tau_4\propto \xi_4^z$}
		\\
		\ \ $A/\sigma$\ \ 
		& \multicolumn{1}{c}{$\rhoc \sigma^2$}
		& \multicolumn{1}{c|}{$b$}
		& \multicolumn{1}{c|}{$\tilde{\gamma}$}
		& \multicolumn{1}{c}{$z$}
		\\
		\hline
		0.01
		& 1.62, 0.03
		& 0.8, 0.4
		& 1.62, 0.21
		& 2.01, 0.22
		\\
		0.014
		& 1.56, 0.02
		& 0.70, 0.15
		& 1.73, 0.09
		& 2.17, 0.11
		\\
		0.02
		& 1.52, 0.01
		& 0.74, 0.14
		& 1.76, 0.15
		& 2.02, 0.12
		\\
		0.03
		& 1.51, 0.01
		& 0.74, 0.05
		& 1.79, 0.16
		& 2.01, 0.12
		\\
		0.04
		& 1.55, 0.03
		& 1.3, 0.4
		& 1.8, 0.4
		& 1.84, 0.26
		\\
		\hline\hline
	\end{tabular}
	\caption{\label{tab:c-exp}
		Critical density $\rhoc$, parameter $b$, and critical exponents $\tilde{\gamma}$ and $z$ of the fluid-tetratic transition
		for several amplitudes $A$ obtained from fitting \eqs{eq:xi-fit}, \eqref{eq:chi-fit}, and \eqref{eq:tau-fit}.
		Simulations at $A=0.03\sigma$ were performed with box size $L=120\sigma$, simulations at other $A$ were performed with $L=80\sigma$.
	}
\end{table*}

\begin{figure}
	(a)\raisebox{\glueexpr\baselineskip -1\glueexpr\height}{\includegraphics[width=\colwidth\textwidth]{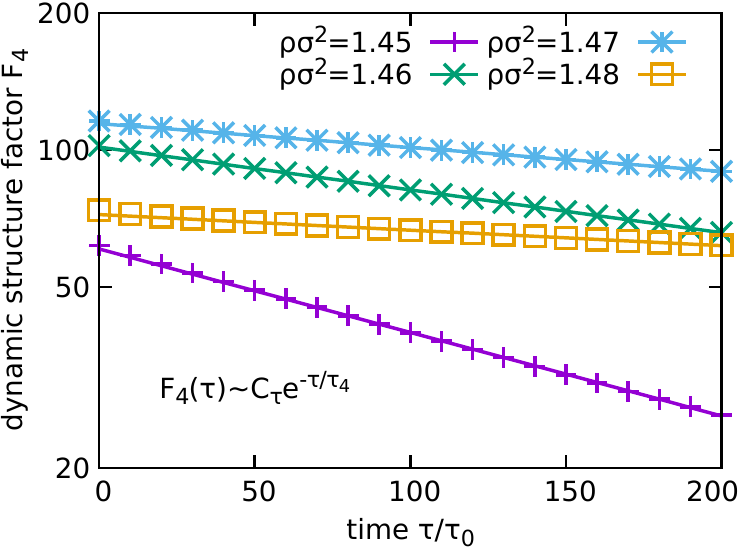}}\\
	(b)\raisebox{\glueexpr\baselineskip -1\glueexpr\height}{\includegraphics[width=\colwidth\textwidth]{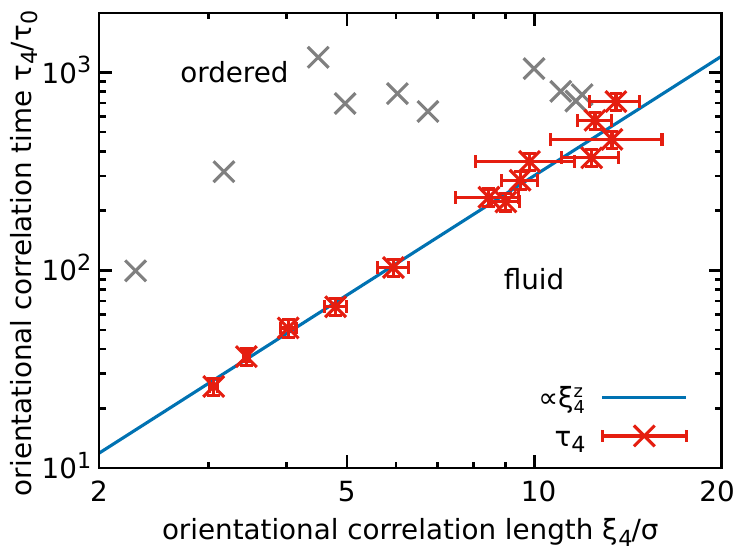}}
	\caption{\label{fig:dyn-corr}(color online)
		(a) Intermediate scattering function $F_4$ as a function of time difference $\tau$
			at fixed wave number $k\sigma =0.052$ obtained at several densities $\rho$ as indicated, amplitude $A=0.03\sigma$, and box size $L=120\sigma$.
			Symbols show simulation data, lines show fits to \eq{eq:F-fit}.
		(b) Correlation time $\tau_4$ as a function of orientational correlation length $\xi_4$
			obtained from the fits of $F_4(k,\tau)$ \bracket{see (a)\,}.
			The line is a fit to \eq{eq:tau-fit} using data of the fluid regime (red symbols with error bars).
			Gray symbols without error bars are ordered states \bracket{cf. \subfig{fig:stat-corr}{b}}.
	}
\end{figure}

The critical slowing down of large patches is quantified by measuring the correlation time $\tau_4$ and the dynamic critical exponent $z$ in the limit of low $k$.
After an initial decay of all but the slowest mode, the long-time asymptotics of the intermediate scattering function $F_4$ has an exponential tail,
\begin{equation}
    \label{eq:F-fit}F_4(k\rightarrow 0,\tau) = C_\tau \exp(-\tau/\tau_4),
\end{equation}
with the constant prefactor $C_\tau < \suscep$ due to the initial decay.
The simulation results and fits are displayed in \subfig{fig:dyn-corr}{a} exemplarily at $A=0.03\sigma$ for $L=120\sigma$.
Again, we observe the same qualitative behavior at all investigated $A$.
To extract the asymptotic exponential decay, we fit data for $\tau\geq50\tau_0$.
The results for $\tau_4$ are shown in \subfig{fig:dyn-corr}{b} as functions of $\xi_4$.
Again, $\tau_4$ diverges with $\xi_4$ as
\begin{equation}
	\label{eq:tau-fit}\tau_4\propto\xi_4^{z},
\end{equation}
which defines $z$ (also shown in \ourOnlyTable).
Assuming that the values are constant along the critical line, the best estimates averaging over the values for different $A$ are
\begin{equation}
	\tilde{\gamma} = 1.73\pm 0.07\ ,\ z = 2.05\pm 0.06.
\end{equation}

\begin{figure}
	\includegraphics[width=\colwidth\textwidth]{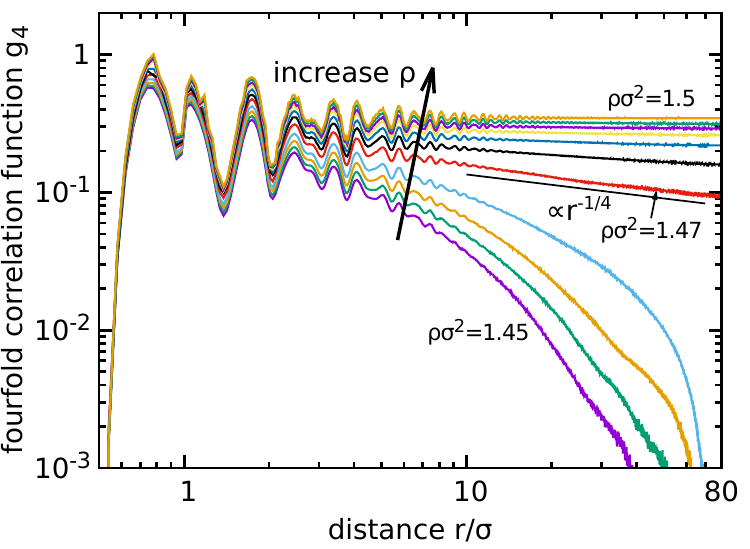}
	\caption{\label{fig:g4}(color online)
		Fourfold correlation function $g_4$ as a function of distance $r$ on a double-logarithmic scale
		at densities ranging from $\rho\sigma^2=1.45$ to $1.5$, amplitude $A=0.03\sigma$, and box size $L=160\sigma$.
		The black straight line corresponding to a power law with exponent $-1/4$ is a guide to the eye.
	}
\end{figure}

Finally, the fourfold correlation function $g_4(r)$ is examined
to estimate the anomalous dimension $\eta_4$ of orientational order.
We define
\begin{equation}
	\begin{split}
		g_4(\rvec)\equiv\frac{1}{\rho N}\left<\sum_{m=1}^{N}\sum_{n=1}^{N}\right.&\delta[\rvec+\rvec_m(t)-\rvec_n(t)]\\
		&\times\psi_4^{(m)}(t)\bar{\psi}_4^{(n)}(t)\left.\vphantom{\sum_{m,n}^{N}}\right>,
	\end{split}
\end{equation}
which is in practice calculated via backwards Fourier transforming $S_4(\kvec)$.
Plotting $g_4$ on a double-logarithmic scale (see \fig{fig:g4}) for $L=160\sigma$, one can distinguish between short-ranged exponential decay for $\rho\sigma^2\leq 1.465$ and quasi-long-ranged algebraic decay $g_4\propto r^{-\eta_{4}}$ for $\rho\sigma^2\geq 1.47$. The algebraic decay at the transition is well described by a power law with
\begin{equation}
\eta_{4}\approx 1/4.
\end{equation}

\begin{figure}
	(a)\raisebox{\glueexpr\baselineskip -1\glueexpr\height}{\includegraphics[width=\colwidth\textwidth]{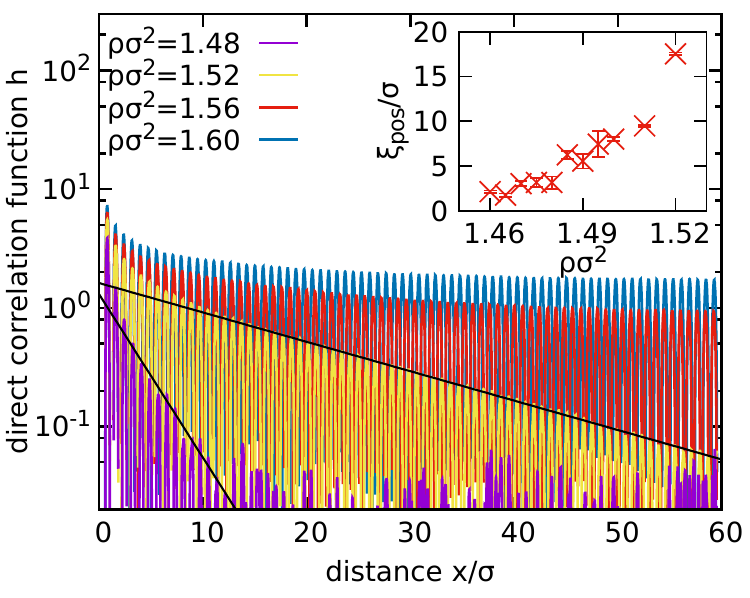}}
	(b)\raisebox{\glueexpr\baselineskip -1\glueexpr\height}{\includegraphics[width=\colwidth\textwidth]{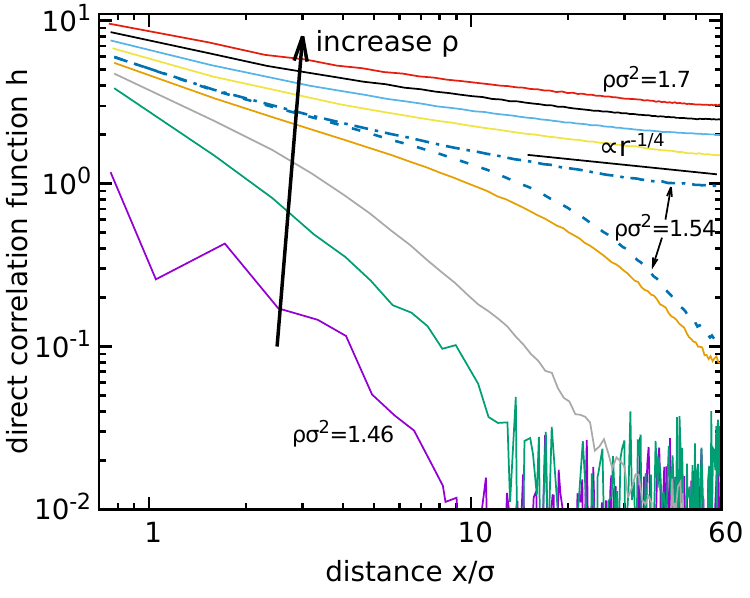}}
	\caption{\label{fig:g-coh}(color online)
		(a) Scan through the \ac{2d} direct correlation function $h$
			along the direction $x$ of the order parameter $\psiglob$ at some exemplary densities as indicated and box size $L=120\sigma$.
			Multiple samples have been aligned and averaged over coherently, see main text.
			The black straight lines show exponential fits of the maxima of $h(x)$ to \eq{eq:rdf-exp}.
			The inset shows the resulting fit values of the positional correlation length $\xi_\mathrm{pos}$ for all obtained exponentials.
		(b) Maxima of the data shown in panel (a) on a double-logarithmic scale
			for a wider range of densities from $\rho\sigma^2=1.46$ to $1.70$.
			For $\rho\sigma^2 = 1.54$ two curves are shown to illustrate non-ergodicity: the dashed and dashed-dotted lines were obtained from tetratic and square lattice initial configurations, respectively. Data shown for higher $\rho$ were obtained with square lattice initial configurations.
			The black straight line (power law $\sim r^{-1/4}$) is a guide to the eye, where the exponent $1/4$ is the upper bound for $\eta_\mathrm{pos}$ predicted by \ac{bkthny} theory~\cite{etapos}.
	}
\end{figure}

Now we turn to the tetratic-square transition, at which the density of free dislocations vanishes.
The length scale associated with this density can be extracted from the decay of the \ac{2d} pair correlation function $g(x,y)$ towards unity.
To reduce noise, multiple samples are averaged coherently~\cite{BernardPRL2011,KapferPRL2015}, \ie,
they are aligned such that their individual global orientational order parameter $\psiglob$ is real and positive.
(Note that this procedure is necessary rather than investigating the radial distribution function $g(r)$, where the azimuth $\phi$ has been averaged over.
Rapid decay of $g(r)-1$ would be insufficient to demonstrate short-ranged positional order, as $g(r)-1$ decays rapidly even in a solid.
The same is true for a naive average, where multiple configurations with different $\psiglob$ orientations are averaged over incoherently.)

\subfiglong{fig:g-coh}{a} shows the (coherent) direct correlation function $h(x)\equiv g(x,0)-1$ at some exemplary densities for $L=120\sigma$.
The system exhibits short-ranged exponential decay of the envelope,
\begin{equation}
\label{eq:rdf-exp}h(x)\propto \exp(-x/\xi_\mathrm{pos})\hspace{0.5cm}\mathrm{for}\hspace{0.5cm}\rho\sigma^2\leq 1.52,
\end{equation}
where $\xi_\mathrm{pos}$ is the positional correlation length, \ie, the typical distance of dislocations.
The inset shows the fitted values of $\xi_\mathrm{pos}$ for all obtained exponentials.
With increasing $\rho$, $\xi_\mathrm{pos}$ increases implying a decrease of the number of dislocations.
Ultimately the system crosses over to algebraic quasi-long-ranged behavior,
\begin{equation}
\hspace{0.5cm}h(x)\propto x^{-\eta_\mathrm{pos}}\hspace{1.0cm}\mathrm{for}\hspace{0.5cm}\rho\sigma^2 > 1.52,
\end{equation}
with $\eta_\mathrm{pos}$ being the anomalous dimension of positional order.
The two types of asymptotics are distinguished by plotting $h(x)$ on a double-logarithmic scale as done for the envelope in \subfig{fig:g-coh}{b}.
At the crossover density $\rho\sigma^2=1.52$ the average number of dislocations in the simulation box is of the order of 1.
Note that for $\rho\sigma^2 > 1.52$ simulations are not ergodic any more at timescales $\sim 10^4 \tau_0$ [as indicated by the two curves shown for $\rho\sigma^2 = 1.54$ in Fig. 7(b) corresponding to tetratic and square initial states, respectively].
Averaging over times with quasi-long-ranged behavior \bracket{dash-dotted curve in \subfig{fig:g-coh}{b}},
the asymptotic decay of $h(x)$ is characterized by an exponent close to $1/4$.
The critical density $\rhopos$ where dislocations become infinitely sparse is slightly higher than the crossover density.
Assuming the same magnitude of finite size corrections for $\xi_\mathrm{pos}$ as for $\xi_4$ at the fluid-tetratic transition we estimate $\rhopos\sigma^2\approx 1.55$.

\section{Directed motion of threefold clusters}\label{sub:facet}

\begin{figure}
	\begin{overpic}[width=0.225\textwidth]{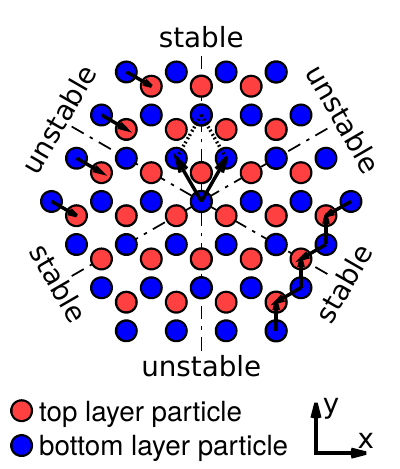}
		\put (-0.1,90) {(a)}
	\end{overpic}\ \ 
	\begin{overpic}[width=0.241\textwidth]{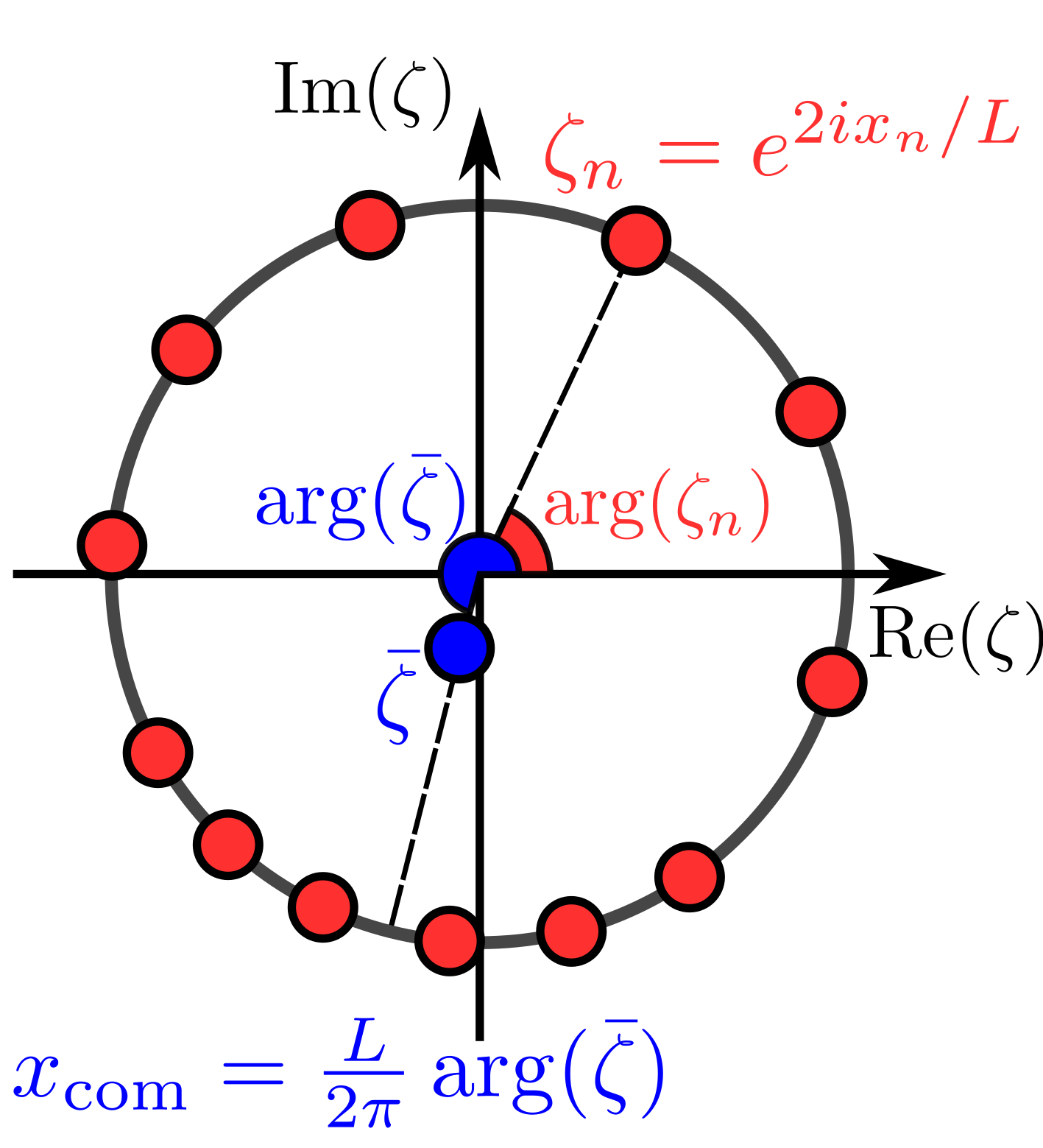}
		\put (-0.1,90) {(b)}
	\end{overpic}
	\caption{\label{fig:lattice}(color online)
		(a) Top view sketch of the threefold lattice consisting of two hexagonal layers of particles (not to scale).
		The rhombus in the center represents a unit cell of the lattice with its long diagonal being one of the three symmetry axes (dashed-dotted lines).
		The orientation of a cluster is measured by the angle $\varphi$ between this diagonal and the $y$ axis of the box.
		Arrows at the edge show how top layer particles are supported by bottom layer particles from the exterior at the stable and unstable facets, respectively, and hence illustrate the asymmetry between these facets.
		In the lattice displayed here, $\varphi=0\degree$.
		This orientation has two equivalent left and right interfaces in configurations with interfaces parallel to the $y$ direction \bracket{as in~\subfig{fig:pd}{\band}}, whereas a lattice rotated by $\varphi=30\degree$ would constitute the maximally asymmetric case with a stable facet to the right and an unstable facet to the left.
		The case $\varphi=60\degree$ is the same as $\varphi=0\degree$ mirror-inverted in $y$ direction and therefore again symmetric with respect to the $y$ axis.
		Therefore, it is sufficient to consider angles between $0\degree$ and $30\degree$.
		(b) Sketch of the calculation of the $x$ coordinate of the center of mass \com{x}
		with periodic boundary conditions via mapping of the $x$ coordinates of the particles onto the unit circle in the complex plane.
		Particles are displayed as red (light gray) circles; the average of the mapped coordinate is depicted as blue (dark gray) circle.
		Symbols are declared in the text.
	}
\end{figure}

\begin{figure*}
	(a)\raisebox{\glueexpr\baselineskip -1\glueexpr\height}{\includegraphics[width=\colwidth\textwidth]{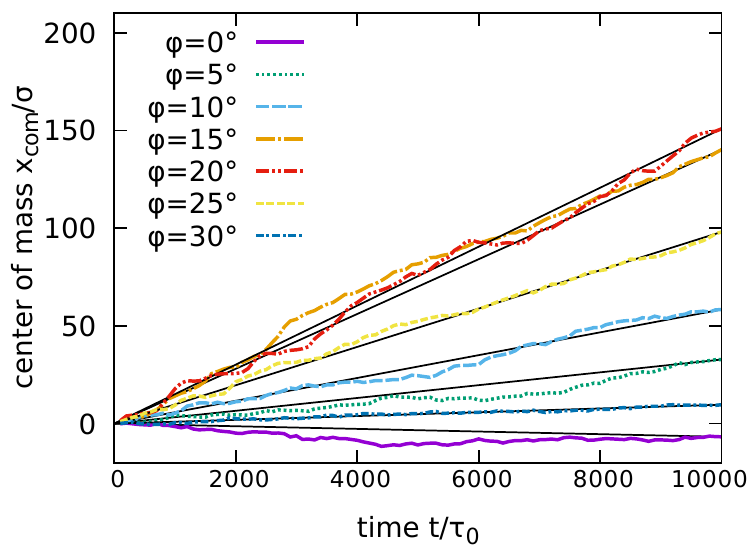}}
	(c)\raisebox{\glueexpr\baselineskip -1\glueexpr\height}{\includegraphics[width=\colwidth\textwidth]{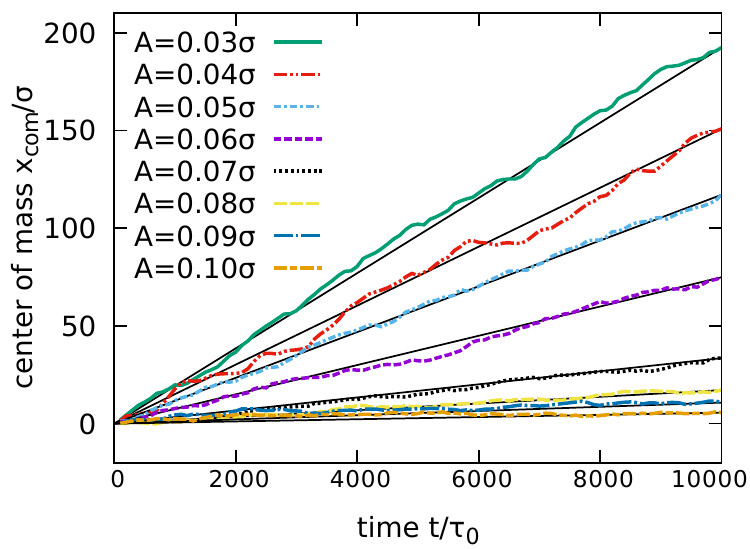}}\\
	(b)\raisebox{\glueexpr\baselineskip -1\glueexpr\height}{\includegraphics[width=\colwidth\textwidth]{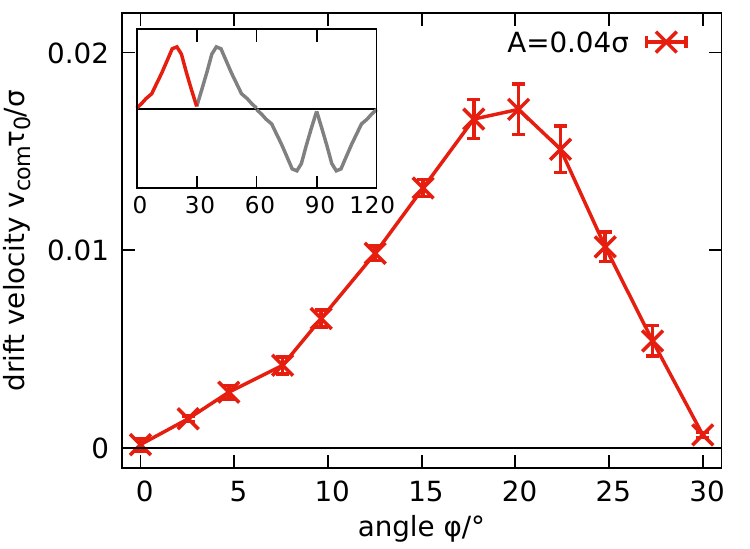}}
	(d)\raisebox{\glueexpr\baselineskip -1\glueexpr\height}{\includegraphics[width=\colwidth\textwidth]{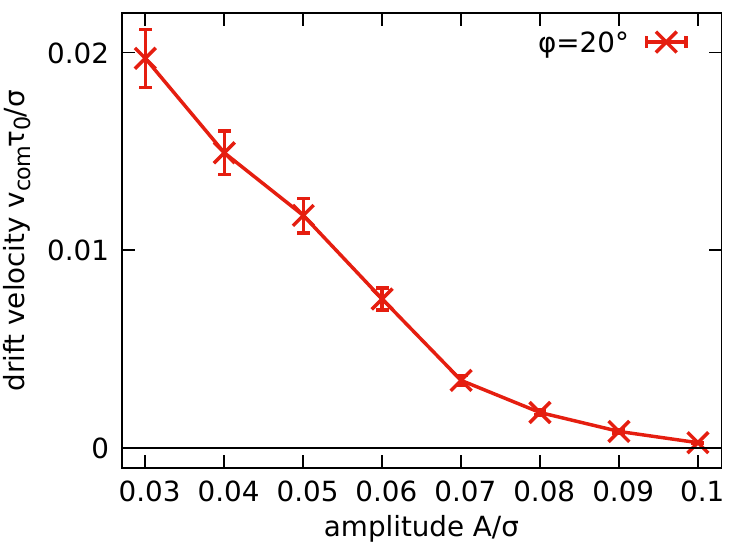}}
	\caption{\label{fig:drift}(color online)
		(a) $x$ coordinate of the center of mass \com{x} as a function of time $t$
		for individual runs at density $\rho\sigma^2=1.7$, amplitude $A=0.04\sigma$, and different cluster orientations $\varphi$ as indicated.
		Black straight lines show end-to-end motion at each $\varphi$ as used to extract the average velocity.
		(b) Cluster speed \com{v} of the center of mass as a function of angle $\varphi$
		obtained from the end to end distance of motion as displayed in (a) and averaged over four individual runs each.
		Lines are guides to the eye.
		By symmetry arguments, it is sufficient to consider the range $0\degree\leq\varphi\leq 30\degree$, see inset.
		(c) Same as (a) but for fixed angle $\varphi=20\degree$ and different $A$ as indicated.
		(d) Same as main panel of (b) but for the center of mass motions displayed in (c).
	}
\end{figure*}

Stripe-shaped threefold clusters like the one shown in \subfig{fig:pd}{\band}
form from compact clusters at sufficiently high $\rho$.
On the other hand, at $\rho$ near the evaporation density, simulation runs initialized with stripes are unstable with respect to compact clusters.

The stripe structure is assumed to occur only with periodic boundary conditions.
With reflecting boundary conditions or in infinite systems, compact cluster shapes would be preferable.
Nevertheless, it is useful to study this topology, as it allows precise control of the fluid-threefold interfaces:
they are straightened to exclude curvature effects and their length is set by the box dimension.
Even the orientation of the lattice grid lines of the cluster is locked via the periodic boundaries and is therefore constant during each simulation run.

To shed light on the mechanism of particle attachment at the interfaces and implications for the shape of compact clusters,
we initialize configurations with percolating clusters in $y$ direction \bracket{as in \subfig{fig:pd}{\band}}. (In principle these clusters could also percolate the box in $x$ direction.)
Box dimensions are slightly adjusted for the unit cell of the cluster to fit in.
Antipodal facets of the threefold cluster phase are inequivalent, as illustrated in \subfig{fig:lattice}{a}.
Top layer particles (red/light gray) at stable facets are supported by two bottom layer particles (blue/dark gray) from the exterior.
A top layer particle at an unstable facet, however, is supported only by a single bottom layer particle.

We initialize clusters with different orientations $\varphi$ \bracket{as depicted in \subfig{fig:pd}{\band}} of the lattice structure with respect to the interface and calculate their drift speed as a measure for differences in the stabilities of the interfaces.
With the threefold cluster being denser than the fluid, the drift of the cluster can be tracked through the $x$ coordinate of the center of mass \com{x}.
As the system has periodic boundary conditions, we identify \com{x} by mapping the $x$ axis to the unit circle as described in the following and illustrated in  \subfig{fig:lattice}{b}.
First, the $x$ coordinates $x_n$ of the particles are mapped to complex numbers $\zeta_n=\mathrm e^{2\pi \mathrm i x_n/L}$.
These are averaged in the complex plane via $\bar{\zeta}=\frac{1}{N}\sum_{n=1}^N \zeta_n$ and the complex phase of the average is mapped back to obtain $\com{x}=\frac{L}{2\pi} \arg(\bar{\zeta})$.

Note that the motion of \com{x} indicates density waves through the periodic copies of the box, even though there is no net in-plane momentum.
We initialize the simulation with zero total momentum and the dynamics conserve momentum in $x$ and $y$ directions.
Indeed, the mean velocity of the particles due to rounding errors in the simulations is lower than $10^{-10}\sigma/\tau_0$ throughout the simulations and is therefore not accountable for the measured directed cluster motion.

The center-of-mass motions for different cluster orientations $0\degree\leq \varphi\leq 30\degree$ are shown in \subfig{fig:drift}{a}.
From these the average cluster speeds are calculated by measuring the end-to-end distance and dividing by the time interval [see \subfig{fig:drift}{b}].
As expected, one observes zero cluster speed for symmetric interfaces, \ie, when left and right interface both are half-way between the stable and unstable facets.
For nonzero angles, the cluster starts to move in the direction where the interface is composed predominantly of the stable facet.
The cluster speed increases with increasing asymmetry between the interfaces.
For angles greater than $20\degree$ the speed decreases again, although the cluster becomes even more asymmetric here.

The mechanism of the advancing cluster relies on particle exchange with wetting films of square symmetry on its boundary.
Particles are absorbed in a zipper-like fashion at the front interface and detach from the rear interface.
The effect of this process is a translation of the cluster in positive $x$ direction
(see movie no.~4 in the Supplemental Material~\cite{Supp}).
When approaching $\varphi=30\degree$, however, the right interface becomes parallel to the stable facet and lacks the kinks necessary for the zipper mechanism to function.
This explains the decrease of the cluster speed at these angles.

We also measure the cluster speed as a function of $A$ as shown in \subfigs{fig:drift}{c} and \ref{fig:drift}(d).
Here one can see that the cluster speed decreases with increasing $A$.
This is paralleled by the clusters becoming more circular in the nonpercolated configurations \bracket{cf.\ \subfigs{fig:pd}{\round}, \ref{fig:pd}(\triang), and \ref{fig:pd}(\bulky)} and hence confirms that the mechanism that leads to the faceted cluster shape is the same as the one driving the cluster propulsion.
Two reasons for the decreased cluster speed are identified.
First, the square film is thinner at higher $A$ and therefore, the zipper mechanism does not work as efficiently.
Second, the decreasing influence of gravity as compared to the driving reduces the asymmetry between the two types of facets.

\section{Discussion and Conclusion}\label{sec:con}

The phase diagram of the vibrated quasi-\ac{2d} granular sphere system exhibits a first-order transition to a threefold lattice as well as a continuous fluid-square transition with intermediate tetratic phase.
The densities on the fluid and threefold side of the first-order transition are consistent with values reported for simulations at $A=0.15\sigma$ by Melby \textit{et al.}~\cite{MelbyJPCM2005}.
The critical amplitude for the inelastic collapse -- best compared in terms of the dimensionless acceleration $\Gamma=A\omega^2/g=1.01\pm 0.14$ -- is also in fair agreement with previous studies (see, e.\,g.,~\cite{OlafsenPRL1998,NieEPL2000}).
All of these previous studies involve tangential friction.
Therefore, the consistency with our results demonstrates that tangential friction is not essential for the phase behavior in fluidized granulates.

The fourfold ordering transition found in experiments~\cite{CastilloPRL2012,CastilloPRE2015} and simulations~\cite{GuzmanPRE2018} is consistent with the topology of the phase diagram presented here.
Quantitative agreement between simulation and experiment is not
expected due to the subtle role of roughness of particle surfaces~\cite{GuzmanPRE2018}.
Our study departs from the earlier studies in two ways.
First, instead of the driving amplitude $A$, we control the transition via the global density $\rho$;
second, the earlier studies examine configurations with a fluid-fluid phase separation
in which the denser of the two fluids undergoes an ordering transition by increasing $A$.
Due to the phase separation, the density of the fluid is not strictly fixed either.
For our set of parameters, there is no stable fluid-fluid phase separation,
which considerably simplifies the analysis
and permits direct control of the density.
Therefore, our approach yields the critical exponents associated with the control parameter $\rho$ without admixture of $A$.
We expect that the tetratic phase is also observable in experiment.
The positional correlation length is $\sim 3$ particles when the tetratic first forms, which indicates that short-ranged order will be detectable even in very small systems.

For the fluid-tetratic transition, precise values for $b$ and $\rhoc$ were calculated.
We precisely measured the critical exponents $\tilde{\gamma} = 1.73\pm 0.07$ and $z = 2.05\pm 0.06$, and find $\eta_4 \approx 1/4$.
At a higher density $\rhopos$, there is a tetratic-solid transition which is also of Berezinskii-Kosterlitz-Thouless type.
Surprisingly, all of the critical properties we measure are consistent with \textit{equilibrium} \ac{bkthny} theory for two-step melting in two dimensions,
which predicts critical exponents of $\tilde{\gamma} = 7/4$, $\eta_4=1/4$~\cite{KosterlitzJPC1974}, and $z = 2$~\cite{BerezinskiiJETP1971}.
The observed asymptotic behavior at melting is consistent with the \ac{bkthny} bound for square lattices of $\eta_\mathrm{pos}\leq 1/4$~\cite{etapos}.
All critical properties are thus remarkably close to the equilibrium predictions, despite the strong driving and irreversibility.

The phenomenological resemblance of the phase behavior to equilibrium systems appears to call for a thermodynamic description
by an extension of equilibrium statistical mechanics to \acp{ness}.
However, there are also important deviations from equilibrium behavior.
In the moderate-amplitude regime, the evaporation density of the threefold cluster is larger than the density of the fluid coexisting with it.
A detailed investigation of coexistence density and evaporation density as a function of system size could clarify the origin of this anomaly.
If the dilution is found also in the infinite size limit, this would imply
that pressure is nonmonotonic as a function of $\rho$,
which is at odds with the postulates of equilibrium statistical mechanics.
Therefore, the dilution would be a nonequilibrium effect caused by the persistent energy flows and dissipation.
Similar effects have been observed for active Brownian particles with hydrodynamic interactions~\cite{BlaschkeSM2016}.
We leave this point to a future study.

As a final nonequilibrium effect, traveling density waves and nontrivial particle currents associated with directed motion of stripe-shaped threefold clusters were found.
The same mechanism is responsible for the faceting of the freestanding threefold cluster at low $A$.
These freestanding clusters, however, do not exhibit directed motion and drift only diffusively.
Future studies should clarify the underlying symmetry breaking in the microscopic dynamics and characterize the resulting energy and particle currents.
The absence of rotational degrees of freedom makes this model a good starting point for tracking energy flows.

\begin{acknowledgments}
TS was supported by the Deutsche Forschungsgemeinschaft as part of the Forschergruppe GPSRS under Grant No.\ ME1361/13-2.
We are grateful to Marcus Bannerman for providing the DynamO \ac{md} simulation program and for advice concerning the adjustment of the code to our needs,
to Uwe T\"auber for helpful comments on an earlier version of the article, and to David Nelson for pointing us to Ref.~\cite{OstlundPRB1981}.
We thank Klaus Mecke for useful discussions and continued support.
\end{acknowledgments}

\bibliography{literature,local}

\end{document}